\documentclass[%
twocolumn,
showpacs,
nofootinbib, nobibnotes,
amsmath,amssymb, aps,
pra,
floatfix,
]{revtex4}
\usepackage{graphicx}
\usepackage{placeins}
\usepackage{dcolumn}
\usepackage{bm}
\usepackage{hyperref}
\usepackage{natbib}
\usepackage{subfigure}
\usepackage{color}

\begin{document}

\preprint{APS/123-QED}

\title{Molecular Piston Engine with Spiking Transduction of Back-action in Hybrid Opto-mechanical Brownian Motors}

%

\author{Yusuf G\"ul}
\email{yusufgul.josephrose@gmail.com.}\affiliation{Department of Physics, Bo\u{g}azi\c{c}i
University 34342 Bebek, Istanbul, Turkey} \affiliation{Department of Physics, Middle East Technical University, TR-06531, Ankara, Turkey} \affiliation{veri-zeka.com}

%
\date{\today}

\begin{abstract}
We consider the Molecular Opto-mechanical systems in back-action amplification of single molecule Raman imaging. Surface Enhanced Raman Scattering (SERS) is mapped into the dissipative cavity optomechanics system of coupled resonators. We investigate the plasmon molecular vibration interactions in strong coupling regimes of cavity opto-mechanics in the presence of impurities. Eigenfunction spectrum is analyzed for the normal mode splitting of photonic and mechanical hybrid system. Back-action transduction of photons into mechanical modes is investigated by avoided crossing due to the nonlinear interactions and Casimir forces in the presence of virtual photons and radiation pressure. In Input-output coupled cavity scheme, both cavity and driving fields are analyzed for the absorption and dissipation of heat in weak and strong coupling regimes of the cascaded cavity setup. In terms of the second order coherence functions, thermodynamic work and heat engine is investigated via phonon lasing of mechanical mode in opto-mechanical implementation of Brownian motors by tuning the competing coupling strengths of impurities and nonlinearities.
\end{abstract}



\maketitle
\section{Introduction}

Molecular opto-mechanics  \cite{Roelli2016} gets raise of the enhancement of the dynamical back-action of the
vibrations in Raman imaging which appears in material science \cite{Pettinger2006}, chemistry \cite{Kneipp1997} and biomedical applications \cite{Qian2008}. Based on the SERS and tip-enhanced Raman
scattering (TERS) \cite{Zhang2013}, the interaction between molecular vibration and the plasmons is analyzed by
mapping the problem into the cavity opto-mechanics \cite{Kippenberg2008, Aspelmeyer2014}. Grounding in the plasmonic cavities, Raman scattering of the electric fields are used to obtain the
chemical structure of molecular characteristics in micro domains \cite{Schmidt2016}. Nano stars \cite{Niu2015},
nanoparticle-on-a-mirror morphologies are accounted for the improvement of the chemical
structure analysis and amplification of the molecular vibrations with high resolution \cite{Park2008, Lorenzo2010, Sonntag2014, Lombardi2016}. Optical antennas \cite{Esteban2010}
are used in the enhancement of the chemical fingerprinting by microscopy capabilities of TERS.

Stochastic thermodynamics at small scales brings out the thermal fluctuations in the description of colloidal particle dynamics such as Brownian Carnot engines \cite{Carnot1824, Prigogine2015, Jarzynski1997, Crooks1999, Seifert2012}. Moreover, heat engines appears as the test bed for the investigation of the thermodynamics in micro-scale realization of photon-matter interactions \cite{Scully2002, Linden2010, Horodecki2013, Correa2014}. Among these, opto-mechanical engines are used to investigate the radiation pressure in extracting work from thermal reservoirs with temperature differences \cite{Zhang2014}. In terms of the cavity and mechanical mode,
normal mode splitting is used to investigate the energy conversion between photons and phonons in opto-mechanical setups \cite{Aspelmeyer2009, Lemonde2013}. Together with the photonlike or phononlike nature of the normal modes, transitions between polaritonic branches describe engine operation like Otto cycle \cite{Sonntag2008}. In these systems, black body radiation background plays an important role in extracting work and infusing excess energy into the intracavity field.

In the enhancement of molecular opto-mechanical systems, hybrid opto-mechanical systems \cite{Mari2015, Dechant2015}, mimicking the cavity QED setups for heat engines, contains both virtual photon occupation and dissipative environmental effects in weak and strong photon-atom coupling regimes \cite{Bergenfeld2014,Song2016, Li2016}. Treating the mechanical mode as the probe, virtual photon radiation is used to describe the displacement of the mechanical mode due to the coupling with the ground state of the cavity \cite{Garziano2015, Macri2018}. Displacement of the mechanical probe can be observed by the modulation of the transduction of the virtual photons into mechanical mode in coupled cavity systems containing nonlinearities \cite{Cirio2017}. In weak coupling regime of opto-mechanical system, dressed ground state does not contain the photons and the system is described by the Jaynes-Cummings interactions. Moreover, in strong coupling regime, ground state becomes dressed with the virtual photons and the system is described by going beyond to the Rotating Wave Approximation (RWA) due to the non-conserving interaction terms \cite{Blais2004, Schuster2007, Larson2008, Didier2011}. Effective models \cite{O'Brien1972, O'Brien1983, Dereli2012} are used to describe the coupled resonator systems without RWA in input-output formalism \cite{Gul2016, Gul2018}.

This paper is organized as follows. In Sec.$2$, we introduce the enhancement of SERS in molecular opto-mechanical systems in the presence of impurities and nonlinearities by effective coupled resonator model. Results and conclusions  are introduced in section $3$ and $4$ respectively.

\section{Model} \label{sec:model}

Plasmonic cavities facilitate the enhancement of the chemical structure identification and molecular vibration \cite{Roelli2016, Schmidt2016} in single and cascaded cavity QED framework \cite{Mari2015}. In this scheme, the bosonic description of the interaction between molecular vibration and radiation field is mapped to molecular opto-mechanical system containing nonlinearities. In molecular opto-mechanical systems \cite{Roelli2016}, containing impurities represented as two level system, our model Hamiltonian is $(\hbar=1)$
\begin{eqnarray}
H&=&\frac{\omega}{2}\sigma_z+\omega_{1}
a^{\dag}_{1}a_{1}++\omega_{2}
a^{\dag}_{2}a_{2} \nonumber\\
&+&[\lambda_{1}(a_{1}+a_{1}^{\dag})
+\lambda_{2}(a_{2}+a_{2}^{\dag})]\sigma_{x}
\end{eqnarray}
\begin{eqnarray}
H_{int}&=&Ja^{\dag}_{1}a_{1}(a_{2}+a_{2}^{\dag})
\end{eqnarray}
where
 $\omega$ is the frequency of the two level system and $\sigma_{x},\sigma_{z}$ are the Pauli operators. Hybrid cavity and mechanical system are described by  the annihilation and creation
operators $a_{1,2}(a^{\dag}_{1,2})$ of electromagnetic field of cavity with frequency $\omega_{1}$ and mechanical oscillator with frequency $\omega_{2}$. Interaction between cavity field and mechanical oscillator is modulated by the parameter $J$ and two level system are represented by the qubit which is simultaneously coupled to cavity and mechanical modes with strengths $\lambda_{1,2}$.

The opto-mechanical interaction is described by the Fabry-Perrot cavity containing qubit interacting with both cavity field and mechanical displacement of mirror coupled to a spring. Besides, nonlinearities occur due to the bouncing off the photons by the mirror and described as the radiation pressure interaction between cavity photon number $a^{\dag}_{1}a_{1}$ and mechanical displacement operator $a_{2}+a_{2}^{\dag}$ with coupling strength $J$ \cite{Cirio2017}.

As a variation of coupled cavity system \cite{Mari2015}, the cascaded cavity scheme makes it possible to describe piston engine and to investigate the thermodynamic description of the system in input-output formalism by the conversion of the heat into the mechanical work in the presence of the incoherent effects.
When the cavity field frequency is higher than the mechanical resonator and the absorption of the virtual photons of the radiation pressure is taken into account, we employed the two-frequency effective coupled resonator
model\cite{Dereli2012,O'Brien1972,O'Brien1983} and describe our system as
\begin{eqnarray}
H&=& H_{sys}+H_{int},
\end{eqnarray}
where system hamiltonian becomes
\begin{eqnarray}
H_{sys}&=&\frac{\omega}{2}\sigma_z+\omega'\alpha^{\dag}_{2}\alpha_{2}
+\omega_{eff}[\alpha^{\dag}_{1}\alpha_{1}
+k_{eff}(\alpha_{1}+\alpha^{\dag}_{1})\sigma_{z}]\nonumber\\
&+&c_{2}[(\alpha^{\dag}_{1}\alpha_{2}+\alpha_{1}\alpha^{\dag}_{2})+k_{eff}(\alpha^{\dag}_{2}+\alpha_{2})\sigma_{z}],
\end{eqnarray}
and the radiation-pressure interaction is obtained as
\begin{eqnarray}
H_{int}&=&J[(\alpha^{\dag}_{1}\alpha_{2}+\alpha^{\dag}_{2}\alpha_{1})
+(\alpha_{1}+\alpha_{2})^{\dag}(\alpha_{1}^{2}-\alpha_{2}^{2})\nonumber\\
&+&(\alpha_{1}+\alpha_{2})(\alpha_{1}^{2}-\alpha_{2}^{2})^{\dag}],
\end{eqnarray}
in terms of the effective mode frequency
\begin{eqnarray}
\omega_{eff}&=&\frac{\omega_{1}k^{2}_{1}+\omega_{2}k^{2}_{2}}{k_{eff}},
\end{eqnarray}and qubit-resonator coupling strength
\begin{eqnarray}
k^{2}_{eff}&=&k^{2}_{1}+k^{2}_{2}
\end{eqnarray} which is treated as the perturbation.
The frequency of the disadvantaged mode in effective model is obtained as
\begin{eqnarray}
\omega'&=&\frac{\omega_{1}k^{2}_{2}+\omega_{2}k^{2}_{1}}{k_{eff}},
\end{eqnarray}
and the coupling strength
\begin{eqnarray}
c_{2}&=&\frac{\Delta k_{1}k_{2}}{k^{2}_{eff}},
\end{eqnarray} describes interactions between disadvantaged and the privileged mode controlled by the
the frequency difference $\Delta=\omega_{1}-\omega_{2}$.

Our effective model is plausible to investigate the mechanical and thermodynamic properties of the coupled system via QED framework in input-output formalism. Description of the plasmonic cavities in QED framework makes it possible to investigate the interaction between radiation pressure and mechanical mirror for Casimir force and Rabi splitting of the hybrid dressed states in SERS. Coupled resonator effective model makes the radiation pressure coupled to the cavity field as a displacement of mechanical probe to describe the modulation of the virtual photon transduction in strong coupling regime. Besides this, flexibility of treating the coupled cavity system in cascaded scheme with effective models give rise to the analysis of the thermodynamic work extraction and heat transfer for piston engine in input-output formalism of the cavity opto-mechanics.


\section{Results} \label{sec:results}
Controlling the competition between ground state transitions and dissipative environment effects
by tuning the radiation interactions leads to observe the effects of Casimir force and Rabi splitting in eigenvalue spectrum and transduction of virtual photons in a coupled resonator system. For this purpose, we investigate the avoided crossing, normal mode splitting in lowest eigenvalue spectrum and coherence functions of the displacement operators in input-output formalism in the presence of virtual photons.

The eigenvalue spectrum of our system is analyzed in five lowest level for weak and strong coupling regimes.
In Fig.$1.$, we investigated the effect of the hopping parameter $J$ representing the nonlinear interactions for Casimir effect and Rabi splitting by ladder of avoiding crossings of the spectrum in terms of the frequency difference $\Delta.$  Fig.$1.a$ shows the avoided crossing for each level with increasing anticrossing numbers as the frequency difference $\Delta$ increase in weak coupling regime, $k=0.1$ and $J=0.1$. Effect of radiation-pressure interaction in eigenvalue spectrum is seen in number of anticrossing in Fig.$1.b.$ As we increase the coupling strengths of cavity field and mechanical mode, $k=0.1$ and $J=0.1$, number of anticrossing is also increasing and avoided crossing occurs also in smaller $\Delta$ values. In Fig.$1.c$ level splitting increase as we go to the strong regime $k=1.0$ and $J=1.0$.

\begin{figure}[h]
\begin{center}
\subfigure[\hspace{0.001cm}]{\label{fig:1a}
\includegraphics[width=0.4\textwidth]{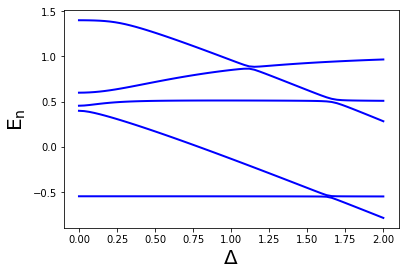}}
\subfigure[\hspace{0.001cm}]{\label{fig:1b}
\includegraphics[width=0.4\textwidth]{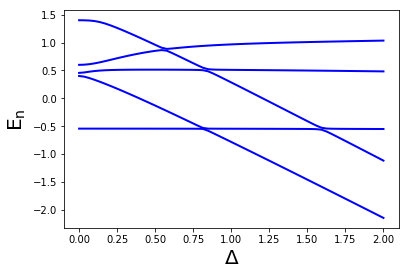}}
\subfigure[\hspace{0.001cm}]{\label{fig:1b}
\includegraphics[width=0.4\textwidth]{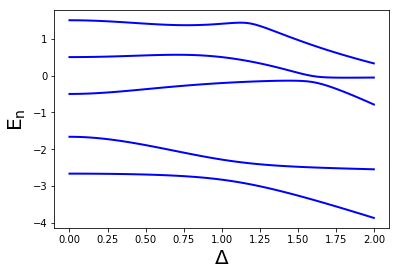}}
\caption{\label{fig3} (Color online) Ladder of avoiding level crossings in eigenvalue spectrum of hybrid photonic and mechanical modes implying the generation of photons from the vacuum. (a) Avoiding crossing in the presence of nonlinear interaction with $k=0.1$ and $J=0.1$.
(b) Effect of increasing the coupling strength by the number of level anticrossing with  $k=0.1$ and $J=1.0$. (c) Level splitting due to the dynamical Casimir Effect in strong coupling regime $k=1.0$ and $J=1.0$. }
\end{center}
\end{figure}

\begin{figure}[!hbt]
\begin{center}
\subfigure[\hspace{0.015cm}]{\label{fig:2a}
\includegraphics[width=0.4\textwidth]{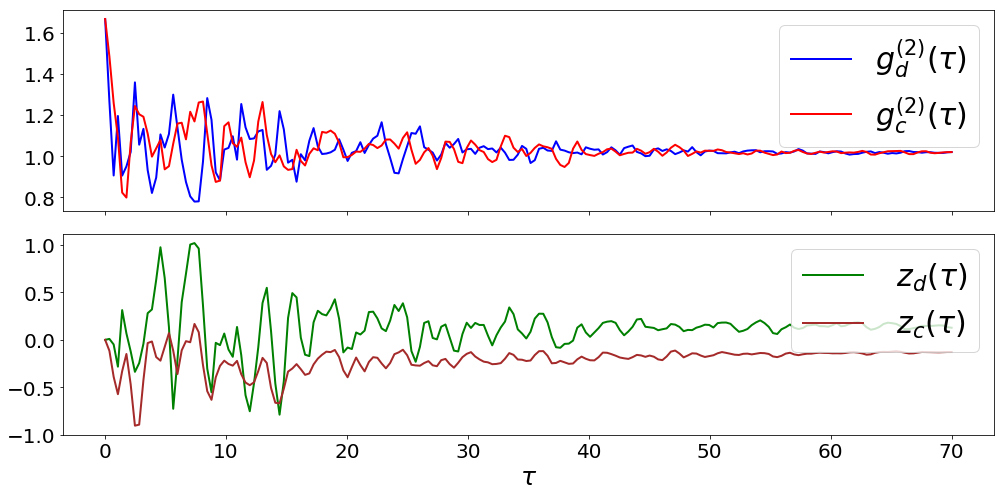}}
\subfigure[\hspace{0.015cm}]{\label{fig:2b}
\includegraphics[width=0.4\textwidth]{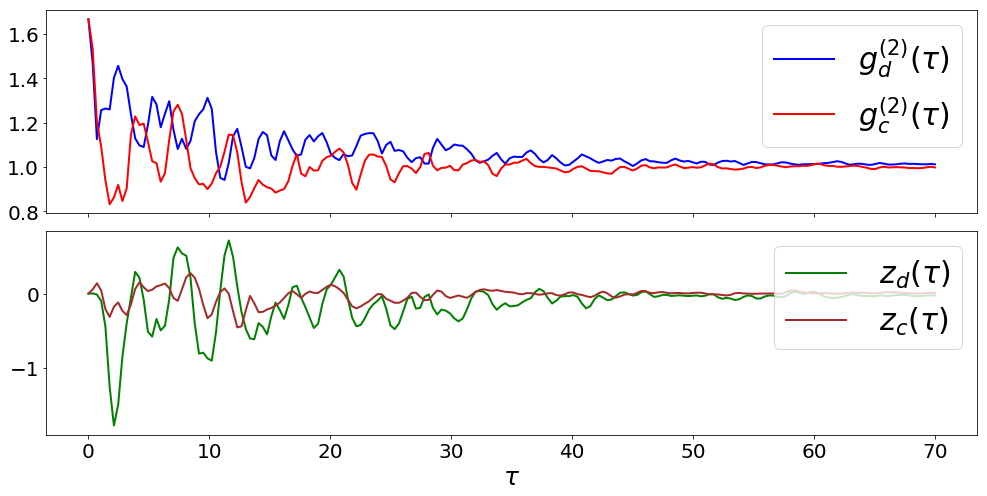}}
\subfigure[\hspace{0.015cm}]{\label{fig:2b}
\includegraphics[width=0.4\textwidth]{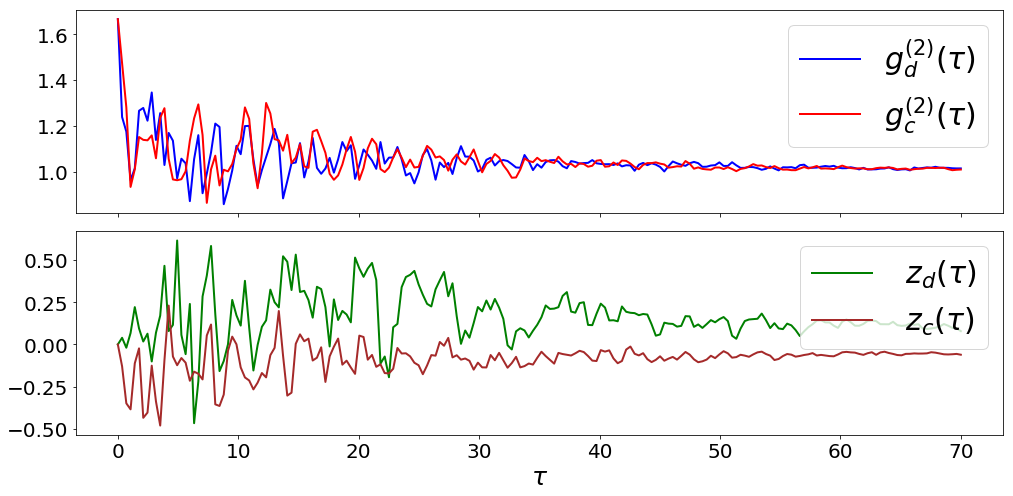}}
\caption{\label{fig3} (Color online) Conversion of heat energy into the coherent mechanical motion and population differences of quadrature operator and normal modes. Convergence of coherence functions $g^{(2)}_{c,d}=1$ implies the phonon lasing of the mechanical mode in weak and strong coupling regimes. (a) Spiking expectation values of population imbalances in weak coupling regime. Conversion of virtual photons in favor of mechanical mode. (b) Effect of decreasing nonlinear coupling strength parameter $J$ on population imbalances. There is no net favor of virtual photon conversion on population imbalances. (c) Virtual photon conversion is in favor of mechanical mode with higher amplitude and spiking form in strong coupling regime  }
\end{center}
\end{figure}

In input-output formalism of cavity opto-mechanics, the competition between virtual photon occupation of the ground state and dissipative environment requires expressing both radiation and mechanical modes in terms of the quadrature operators \cite{Garziano2015, Macri2018, Cirio2017}. For the purposes of the conversion between radiation and mechanical mode, we use the displacement operators
$X_{i}^{+}=\frac{1}{2}(a_{i}+a_{i}^{\dagger})$ where $i=1,2$ represents the resonators in cascaded scheme of opto-mechanical system. Then, in terms of the normal modes $\alpha_{1,2}$ of our effective model \cite{Gul2016, Gul2018}, we use the quadrature operators $X_{1}^{+}=(\alpha_{1}+\alpha_{1}^{\dagger})$ and $X_{2}^{+}=(\alpha_{2}+\alpha_{2}^{\dagger})$ to represents the cavity mode and driving mechanical mode due to the incoherent absorption heat. Conversion of heat from a hot optical/microwave heat bath to mechanical resonator is treated as the thermal motor. Following the idea of production of thermodynamic work of ratchet machine by extracting from random
Brownian noise \cite{Smoluchowski1912, Scovil1959, Feynman1963, Hanggi2009}, our effective model is treated as Brownian motor \cite{Mari2015}. Conversion of virtual photons into mechanical coherent motion is taken into account as an enhancement for the dissipative and noisy opto-mechanical cavity dynamics. Thus, the dissipative environmental effects are analyzed by the master equation in QED framework. Then, governing equation for open dissipative system is written as
\begin{eqnarray}
\frac{d\rho}{dt}=-i[H,\rho]+{\cal L}\rho,
\end{eqnarray}
where the Liouvillian superoperator ${\cal L}$ is described by
\begin{eqnarray}
 {\cal L}\rho&=&\sum_{j=1,2}(1+n_{th})\kappa{\cal D}[\hat\alpha_j]\rho+n_{th}\kappa{\cal D}[\hat\alpha_j^\dag]\rho\nonumber\\
 &+&\gamma{\cal D}[\sigma]\rho+\frac{\gamma_\phi}{2}{\cal D}[\sigma_z]\rho,
 \end{eqnarray}
where $n_{th}$ represents the  average thermal photon number. In our simulations of dissipation effects, taking the value
$n_{th}=0.15$ corresponds to $100$~mK \cite{Dereli2012}.
${\cal D}$ represents the Lindblad type damping superoperators,
$\kappa$ denotes loss rate. Relaxation and
dephasing rates of the qubit are $\gamma$ and $\gamma_\phi$. We take resonator decay parameters
$\kappa_{1}=\kappa_{2}=0.001$ and qubit relaxation and dephasing
parameters $\gamma=0.001, \gamma_{\phi}=0.01$ with the thermal
occupation number $n_{th}=0.15$.

In piston engines, conversion of heat results in the lasing regime of coherent motion of mechanical probe. To investigate the coherent absorption of the heat, we employed
the second order coherence functions of photonic and mechanical mode given by
\begin{eqnarray}
g_{i}^{(2)}=\frac{O^{\dag}_{i}(t)O^{\dag}_{i}(t+\tau)O_{i}(t)O_{i}(t+\tau)}{O_{i}^{\dag}(t)O_{i}(t)}
\end{eqnarray}
where $i=c,d$ are used in place of cavity $X_{c}^{+}=(\alpha_{1}+\alpha_{1}^{\dagger})$ and driving mechanical probe mode $X_{d}^{+}=(\alpha_{2}+\alpha_{2}^{\dagger})$.

We investigate the transduction of the radiation pressure into the mechanical motion in the presence of  virtual photons and multilevel excitations. We use the population difference $z_{c}(\tau)= \langle X^{+}_{1}\rangle - \langle \alpha_{1}\rangle$ and $z_{d}(\tau)= \langle X^{+}_{2}\rangle - \langle \alpha_{2}\rangle$ of each individual cavities in terms of the expectation values of quadrature operators $ X^{+}_{1,2}$ and normal modes $\alpha_{1,2}$.

In Fig.$2$, we analyzed the second order coherence functions $g_{c,d}^{(2)}$ of cavity and driving mechanical mode in weak $k=0.5$ and strong $k=1.0$ coupling regimes.
In each of the figures $2.a, b, c$, both $g_{c,d}^{(2)}$ starts from the thermal states $g^{(2)}_{c,d}\gg1,$ converges to the coherent states $g^{(2)}_{c,d}=1$ indicating the lasing of the coherent mechanical motion. Moreover, both $ z_{d}$ and $ z_{c}$ evolves in spiking form due the incoherent effects of the environments in weak and strong coupling regimes. Fig.$2.a$ shows population imbalance $ z_{d}$ ($ z_{c}$) gets centered around a slightly bit above (below) zero in weak coupling regime $k=0.5$ which implies the nonlinearities are in the favor of the conversion of the heat into the mechanical mode with $J=1.0$ in the presence of virtual photons. Whereas, in Fig.$2.b$, when we decrease the nonlinear interaction strength $k=0.5, J=0.5$, both $ z_{d}$ and $ z_{c}$ centered around zero. Fig.$3.a$ shows that nonlinear coupling strength $J=1.0$ dominates the virtual photon conversion in strong coupling regime $k=1.0$.

Together with the second order coherence functions, population imbalances make our effective model well suited to analyze the enhancement of the SERS in the presence of impurities and nonlinearities. While second order coherence functions are beneficial to investigate the coherent mechanical motions of the probe carrying mirror, population imbalances due to the virtual photons in individual cavities presents the competition between coupling strengths of impurities and nonlinearities in the effective model. This makes our model plausible for adapting to the different schemes of cavity opto-mechanics in engine analysis. Besides this, our effective model is treated  without RWA in weak and strong coupling regimes of cavity QED framework.

\section{CONCLUSION}\label{sec:conclusion}
In conclusion, we investigate the molecular opto-mechanical systems for SERS in QED framework. We construct the hybrid coupled cavity system scheme in the presence of impurities and nonlinearities. In cascaded cavity scheme, we analyzed the transduction of radiation field into the mechanical mode by the conversion of the heat energy into the thermodynamic work. Interaction of cavity field and mechanical probe is studied in Energy eigenvalue spectrum for Casimir forces and Rabi splitting. Employing the effective model beyond the RWA in weak and strong coupling regimes, second order coherence functions are used to see the effect of virtual photon conversion in coherent phonon lasing by tuning the nonlinearities.

\begin{acknowledgements}

Y. G. gratefully acknowledges support by Bo\u{g}azi\c{c}i University
BAP project no $6942$.
\end{acknowledgements}

%

\end{document}